\documentstyle[twocolumn,psfig,fleqn,amstex,rotating]{mn}
\def\etal{{et al.\ }}

\def\lsim{~\rlap{\raise 0.4ex\hbox{$<$}}{\lower 0.7ex\hbox{$\sim$}}~}
\def\gsim{~\rlap{\raise 0.4ex\hbox{$>$}}{\lower 0.7ex\hbox{$\sim$}}~}
\def\dd{{\rm d}}
\def\dl{D_{\rm L}}
\def\dls{D_{\rm LS}}
\def\ds{D_{\rm S}}
\def\jaro{Jaroszy$\acute{\rm n}$ski}
\def\rg{r_{\rm g}}

\def\l0{L_\ast(0)}

\def\s0{S_\ast(0)}

\def\omg0{\Omega_0}
\def\dd{{\mathrm d}}

\def\a2{\alpha^{(2)}}

\def\mpc3{\ {\rm Mpc^{-3}}}
\def\gpc3{\ {\rm Gpc^{-3}}}

\begin{document}

\title[]
{ Monitoring lensed starlight emitted close to the Galactic center}
\author[Nusser \& Broadhurst]{Adi Nusser$^1$ and Tom Broadhurst$^{2}$\\\\
$^1$Physics Department and the Asher Space Researrch Institute, Technion, Haifa 32000, Israel\\
$^2$ School of Physics and Astronomy, Tel-Aviv University, Israel\\}
\maketitle

\begin{abstract}

We describe the feasibility of detecting the gravitational deflection
of light emitted by stars moving under the influence of the massive
object at the Galactic center. Light emitted by a star orbiting
behind the central mass has a smaller impact parameter than the star
itself, and suffers the effect of gravitational lensing, providing a 
closer probe of the central mass distribution and hence a
stricter test of the black hole hypothesis. A mass of $4.3\times
10^{6} M_{\odot}$ causes a $0.1-2\rm mas$ deviation in the apparent
position of orbiting stars projected within $10^{\circ}$ of the
line of sight to the galactic center. In addtion, we may uniquely
constrain the distance to the center of the galaxy because lensing
deflections constrain the ratio $\rg/R_{0}$ of the Schwarzschild
radius to the distance to the black hole, $R_{o}$, whereas the ratio
$\rg/R_{o}^{3}$ is obtained by fitting the orbit.

\end{abstract}



\section {Introduction}
\label{sec:introduction}

Observations of the stellar motions near the Galactic center clearly
reveal the existence of a very massive compact dark onject (Genzel
\etal 2000, Gezari \etal 2002, Figer \etal 2003, Genzel \etal 2003,
Hornstein 2003, Sch\"{o}del \etal 2003, Ghez \etal 2003). The analysis
of Ghez \etal (2003) of the orbital motions of 22 stars yields a
central dark mass of $4\pm 0.3 \times 10^{6}\left(\frac{R_{o}}{8\rm
kpc}\right)^{3}$ where $R_{o}$ is our distance from the galactic
center. This is nearly twice the mass obtained from earlier velocity
dispersion measurements (Genzel \etal 2002, Ghez \etal 2003).  The motion of the
star with the smallest impact parameter indicates that the mass of the
central object is confined to lie within a maximum distance of $90\rm
AU$ from the center, limiting or excluding some of the proposed
alternatives to the black-hole model (Ghez etal 2002).

The deviations from strict Keplerian orbits discussed (Ghez, private
communication) are tantalizing but small and will be very interesting
if they prove significant.  Further monitoring and future higher
resolution measurements, may uncover relativistic effects (\jaro\
1998, Pfahl \& Loeb 2003, Weinberg \etal 2004) but are best explored
with higher resolution observations using space interferometry or a
planned giant telescopes (e.g. the Thirty Meter Telescope, Weinberg
\etal 2004).  Here we consider the signature of light deflection by
gravitational lensing on the trajectories of the stars orbiting the
central massive object. \jaro\ (1998) discussed the statistics of
proper motions in the presence of weak lensing and examined the
apparent distortions of the trajectories only in the strong lensing
limit.  Lensing in the weak limit is much more typical, and here we
examine the deflection of light from stars moving behind the central
mass, particularly in the case of inclined trajectories which are of
large radius but pass close in projection to the central mass, so
that the lens-source separation is larger than the impact parameter
of the light, generating significant light deflection.  We will show
that these effects are comparable or only a little smaller than the
relativistic effects reported by Weinberg \etal (2004) and hence at
the very least represent an important correction for stars orbiting behind
the central mass.
  
  Light reaching us from stars with low inclination angles passes
closer to the central mass than do the stars themselves. Therefore
light deflection effects can probe the inner mass distribution
allowing a stricter test of the black hole hypothesis. Here we study
the way light is deflected with simple constrasting models for
the distribution of matter within the currently unconstrained region
interior to $100\rm AU$ from the center of mass.

\section{The Equations of light deflection}
\label{sec:equations}
Let $\beta$ and $\theta$ be, respectively, the source and image
angular distanes from the center of the lens.  Also let $\dl$ and
$\ds$ be the distances from the observer to the lens and to the
source, respectively. The distance between the lens and the source is
$\dls$.  The light bending equation by a spherical mass distribution
in the thin lens limit is
\begin{equation}
\label{eq:thinlens}
\theta=\beta+\frac{\dls}{\ds}\frac{4 G M(<\xi)}{c^{2 }\xi} \; ,
\end{equation}
where $\xi=\dl \theta $ and 
\begin{equation}
\label{eq:msigma}
M(<\xi)=
\int_{0}^{\xi} 2\pi \xi' \dd \xi'\int_{-\infty}^{\infty}\rho(\xi',z)\dd z \; .
\end{equation}
We will base our results on (\ref{eq:thinlens}) which is valid in the
thin lens limit. This approximation is valid for our the distances 
and mass of relevence here, as can be
seen by comparing the thin lense approximation with the full solution to the geodesic equation.
Using Fermat's principle in a Schwarzschild metric (Landau \& Lifshitz
1975) the path of light is given by
   \begin{equation}
 \label{eq:deflect}
 \frac{\dd^{2} u}{\dd \phi^{2}}+u=\frac{3}{2}\rg u^{2} \; 
 \end{equation}
 where $\rg=2GM/r$ is the Schwarzschild radius, $u\equiv 1/r$, and
 $\phi$ is the angular position in the plane defined by the path. We
 assume the observer to be at $\phi=\pi$.  Given a direction for the
 propagation of light as it leaves the star this equation could be
 solve to obtain $r(\phi)$.  However, we do not know a priori which
 initial direction corresponds to a light ray seen by the observer.
 The initial direction, however, could be varied until the desired
 solution is found.  Instead of this iterative approach, we adopt this
 approximate scheme described in Landau \& Lifshitz (1976, p84). In
 this scheme the desired solution is found as a boundary value problem
 without resorting to iterations. We have tested this scheme in
 several cases and found very close agreement with the solution
 obtain by numerical integration of (\ref{eq:deflect}).

\begin{figure}
\centering
\mbox{\psfig{figure=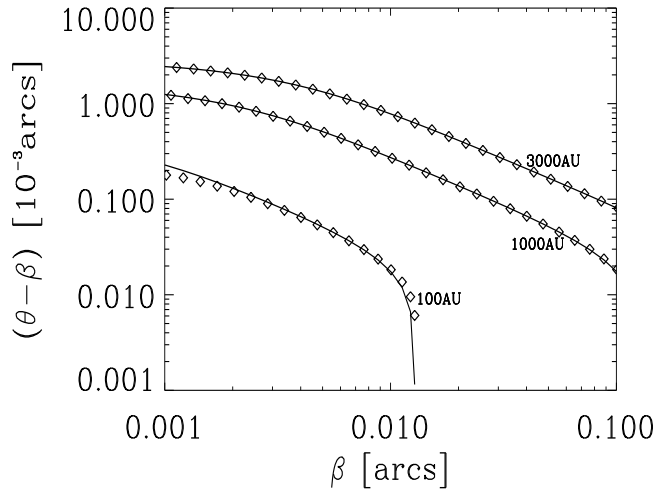,height=3.0in,width=3.5in}}
\caption{
The deflection angle  as a function of the
source angular position on the sky for stars with 
distances of 100, 1000, and 3000AU behind the mass center.
The solid curves and the points correspond, respectively, to
the thin lens approximation and to the solution of the 
geodesic equation}
\label{fig:compare}
\end{figure}

In figure (\ref{fig:compare}) we show plot the deflection angle
$(\theta-\beta)$ obtained from the thin lens approximation (the
solid lines) and the full path equation (the points). The deflections
are computed for sources lying on three edge-on circles centered a
point mass lens. The lens has a mass is $M=4.3\times 10^{6}M_{\odot}$
and $R_{o}=8{\rm kpc}$ is used to the distance to the lens.  The
deflections are computed only for the far side of the circles as the
deflection of light emitted by a source between the lens plane and the
observer is negligible.  This figure demonstrates that the thin lens
approximation is sufficiently accurate when the deflections are larger
than $0.1\rm mas$ even for stars as close as 100AU to the lens.

\section{results}
In addition to light deflection by a point mass (black hole) we
present results for the following mass distributions: a constant
density core, and an isothermal density profile $1/r^{2}$, both
truncated at $r=r_{\rm t}$.  The orbits of stars analyzed by Ghez
\etal (2003) probe the mass distribution outside the current minimum
observed impact paramater of $r=90Au$. These orbits are claimed to be
consistent with Keplerian orbits, so we can assume that all the mass
governing the orbits of stars and so we set $r_{\rm t}=90\rm AU$.  For
the total mass inside $r_{\rm t}$ we take $M=4.3\times
10^{6}M_{\odot}$, corresponding to a Schwarzschild radius of $1.2
\times 10^{7}\rm km$, in the black hole case.  For all the mass
distributions we consider here the integral in (\ref{eq:msigma}) can
be expressed in terms of elementary functions.

\begin{figure}
\centering
\mbox{\psfig{figure=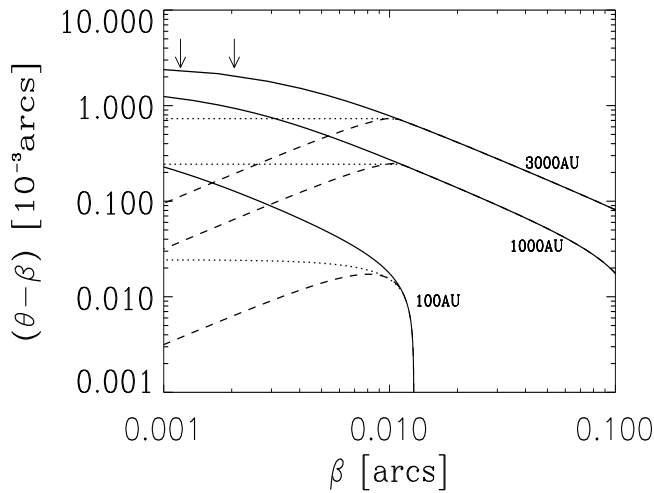,height=3.0in,width=3.5in}}
\caption{The deflection angle  as a function of the
angular position on the sky for edge on circular orbits.
The solid, dotted, and dotted  curves, respectively,  correspond to
lensing by a point mass, a truncated singular isothermal sphere, and
a truncated core of constant density.
 The left and right arrows indicate the angular positions
 corresponding  to the Einstein rings for
 the middle and upper   solid curves (see text), respectively. }
 
\label{fig:one}
\end{figure}

Figure (\ref{fig:one})  shows the
the deflection angle, $\theta-\beta$ as a function of the source position 
$\beta$, for three values of the source distance from 
the center: 100, 1000, and $3000\rm AU$, as indicated in the figure.
We only consider  the weak lensing limit and 
therefore show results only for the bright image. 
The figure shows deflection for the bright image 
The solid, dotted, and dashed curves, respectively,  correspond to deflection 
by a point mass, a  $1/r^{2}$ distribution, and by a constant density.
The right and left arrows indicate the locations of the Einstein 
rings for the 3000AU and 1000AU solid curves, respectively.

\begin{figure}
\centering
\mbox{\psfig{figure=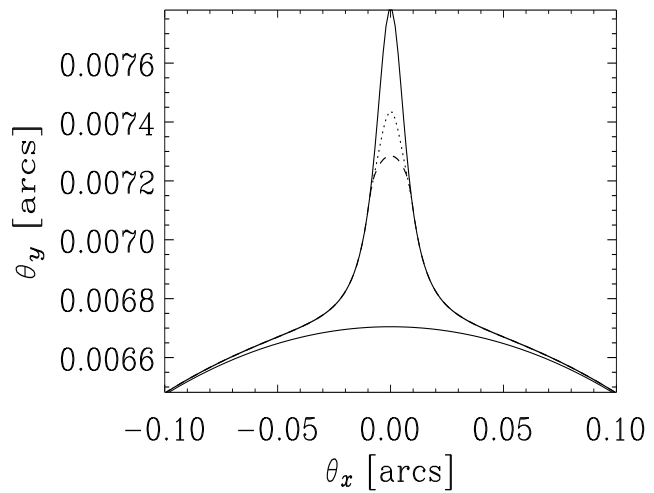,height=3.0in,width=3.5in}}
\caption{The projected trajectories on the sky including the effect
of light bending. For clarity, the deflection angles were
multiplies by a factor of 100. Shown is a segment of the  expected 
trajectories for 
a circular orbit of radius 3000AU and making an angle of $1^{\circ}$ 
with the lone of sight. The solid and dotted curves
correspond,respectiveley, to a point mass and an isothermal sphere truncated at $90$AU. }
\label{fig:two}
\end{figure}

In figure (\ref{fig:two}) we plot the projected trajectories of a
circular orbit as they would be see in the sky.  The radius of the
orbit is 3000AU and the orbital plane has an angle of $1^{\circ}$ with
the l.o.s, The thin solid lines correspond to unlensed
trajectories. The thick solid, dotted, dashed correspond to the a
point mass, $1/r^{2}$ density and constant density, respectively.

These figures demonstrate that the deflections are measurable for a
wide range of $\beta$ for astrometric resolution of $0.1\rm mas$.
Deflections larger than $1 \rm mas$ are restricted to very small
values of $\beta$ corresponding to sources lying within a projected
separation of $300\rm AU$ from the center and are therefore rather
improbable (cf. Ghez \etal 2003).  The deflections by the three mass
distributions differ only for rays passing within $r=r_{\rm t}=90\rm
AU$ of the center, following from our definition.

\section{Discussion} 

We have shown, rather surprisingly, that the gravitational lensing of
stars on orbits about the large mass at the galactic center is a
potentially detectable effect by its perturbation to the observed path
of stars passing behind the central mass. We may expect $\sim 0.1mas$
deflections and these should be measurable with a sufficient number of
accurate observations made with the next generation of space based
interferometers or giant ground based telescopes.  In general the
light from stars orbiting behind the central mass must have a smaller
impact paramter than the actual stellar orbit, by virtue of the
projection, hence gravitational lensing allows the
distribution of the central mass to be explored to within a smaller radius 
may therefore constrain better the black hole hypothesis.  The
lensing effect will be important for orbits whose projected angular
separation is small and in these cases it becomes a significant
correction to the higher order relativitic effects of frame dragging
and orbital precession, for which the anticipated deflections may be
comparable or somewhat larger.
 
 In a coordinate system with an origin at the galactic center, the
apparent deflection of a star lying at an angle $\phi$ with the line
of sight, is $\theta-\beta=\tan (\phi) \dl/(2\rg)$, independent of the
star's distance from the center (see equation \ref{eq:thinlens}).
Substituting $\rg=1.2\times 10^{7}\rm km$, and $\dl=R_{o}=8\rm kpc$
this means that for an astrometric resolution of $\Delta_{0}$ we
should be able to measure deflection of light emitted by all stars
lying in a cone of $11^{\circ}(\Delta_{0}/0.1{\rm mas})$ behind the
galactic center. This corresponding range in the orbital inclination
angle (i.e., the angle between the line of sight and the normal to the
orbital plane) is $85^{\circ}\lsim i \lsim 95^{\circ}$.  The
distortions in the orbits should be detectable by monitoring the stars
around the central dark mass with future telescopes.  According to
Weinberg \etal (2004) the proposed Thirty Meter telescope (TMT) will
be able to measure the orbits of $\sim 100$ stars at distances of
$300-3000\rm AU$ from the galactic center.  The astrometric accuracy
of the TMT will be 0.5mas and could be as good as 0.1mas.  According
to Table 3 of Ghez \etal (2003) the orbits of 2 out of 7 stars have
inclinations in the range $85^{\circ}\lsim i\lsim 95^{\circ}$ range.
Therefore in a sample of a 100 stars we expect more than ten stars to
lie within the range of interest here.

In practice the lensing distortions can be detected by determining the
orbital parameters for each star from the un-lensed majority of the
the orbit.  The deflection can then be detected as the residual
between the parameterized orbit and the lensed part of the orbit.
Given the source's angular position, the deflection are proportional
to $M/R_{o}^{2}$. This differs from the constraints on $M/R_{o}^{3}$ $
M/R_{o}$ derived, respectively, from the projected trajectory and from
radial velocity measurements (Eisenhauer 2003). Note that the
magnification of the emitting stars in the weak lensing limit may rise
to a level of $10^{-3}$ which might seem detectable, however in
practice crowding by fainter stars will limit the accuracy of relative
flux measurements.

General relativistic effects perturbing the physical orbits of stars
have been studied thoroughly by Weinberg \etal (2004). They are
important for stars in the inner central regions and therefore contain
complementary information to the distortions by light deflection. In
terms of the deflection of light, the rotation of a black hole with
angular momentum parameter $a$ (normalized such that $a=\rg/2$ for
maximal rotation) produces a first order correction to
(\ref{eq:thinlens}) for light rays propagating in the equatorial plane
of $\Omega^{3}\rg a /(\rg u -1)^{2}$ added to the left hand side. In
this $\Omega$ is the inverse of the nearest approach distance and
since $a<\rg/2$, this correction is significant only for rays grazing
spheres of radii of a few $\rg$ and so the chances of detecting black
hole spin via lensing of orbiting stars is therefore very small.


\section{Acknowledgment}
This research is supported by a grant from the Asher Space 
Research Institute.




\begin{thebibliography}{}
\bibitem{}Alexander T., Sternberg A., 1999, ApJ, 520, 137
\bibitem{} Bartelmann, M., \& Schneider, P., 1999, astro-ph/9912508
\bibitem{} EisenhauerÊF., Sch$\acute{\rm o}$delÊR., GenzelÊR., OttÊT., TeczaÊM., AbuterÊR., EckartÊA., Alexander T., 2003, ApJ, 597, 121
\bibitem{}FigerÊD.ÊF., GilmoreÊD., KimÊS.ÊS., MorrisÊM., BecklinÊE.ÊE., McLeanÊI.ÊS., GilbertÊA.ÊM., GrahamÊJ. R., LarkinÊJ. E., LevensonÊN.ÊA., TeplitzÊH.ÊI., 2003, ApJ, 599,1139
\bibitem{} Genzel R., Pichon C., Eckart A., Gerhard O.E., Ott T.,
2000, MNRAS, 317, 348
\bibitem{} GenzelÊR., Sch$\acute{\rm o}$delÊR., OttÊT., EisenhauerÊF., HofmannÊR., LehnertÊM., EckartÊA., AlexanderÊT., SternbergÊA., LenzenÊR., ClŽnetÊY., LacombeÊF., RouanÊD., RenziniÊA., Tacconi-GarmanÊL.ÊE., 2003, ApJ, 594, 812
\bibitem{}Ghez A.M., Salim S.,  Hornstein S.D., Tanner A., Morris M.,   Becklin E.E.,
2003, astro-ph/0306130 
\bibitem{}Hornstein S.D., Ghez A.M., Tanner A., Morris M., Becklin E.E., 
Wizinowvich P., 2002, ApJ, 577, L9
\bibitem{} Landau L.D., Lifshitz E.M., 1975, {\it The Classical Theory
of Fields}, Pergamon.
\bibitem{} Landau L.D., Lifshitz E.M., 1976, {\it Mechanics}, Pergamon.
\bibitem{} Jaroszy$\acute{\rm n}$ski
 M., 1998, Acta Astron., 48, 413
\bibitem{}Pfahl E., Loeb A., 2003, astro-ph/0309744
\bibitem{}Sch\"{o}del R., Ott T., Genzel R., Eckart A., Mouawad N., Alexander T.,
2003, ApJ, 596, 1015
\bibitem{} Weinberg N.N., Milosavlejevi{$\acute{\rm c}$} M., Ghez A.M.,
 2004, astro-ph/0404407
 \end{thebibliography}
\end{document}